# Inhomogeneous broadening of optical transitions observed in photoluminescence and modulated reflectance of polar and non-polar InGaN quantum wells

Michał Jarema,[1,*] Marta Gładysiewicz,[1] Łukasz Janicki,[1] Ewelina Zdanowicz,[1] Henryk Turski,[2] Grzegorz Muzioł,[2] Czesław Skierbiszewski,[2] and Robert Kudrawiec[1,**]

[1] *Faculty of Fundamental Problems of Technology, Wrocław University of Science and Technology, Wybrzeże Wyspiańskiego 27, 50-370 Wrocław, Poland*
[2] *Institute of High Pressure Physics, Polish Academy of Sciences, Sokołowska 29/37, 01-142 Warsaw, Poland*

*E-mail: michal.jarema@pwr.edu.pl
**E-mail: robert.kudrawiec@pwr.edu.pl

We analyze the broadening of interband transitions in InGaN/GaN quantum wells (QWs) resulting from structural inhomogeneities. We identify the role of polarization-induced electric field in the mechanism behind the inhomogeneous broadening observed in photoluminescence (PL) and electromodulated reflectance (ER) spectra of InGaN QWs dedicated to green/blue lasers. The spectra of both polar and non-polar QWs are simulated within the random QW model while distinguishing contributions of individual transitions, and are compared for the same magnitudes of QW inhomogeneities (QW width and indium content fluctuations). On this basis we interpret the ER and PL spectra measured for a polar multiple QW sample. The built-in electric field responsible for shifting the emission wavelength to red, enhances at the same time the broadening of optical transitions. It is clearly shown that for polar QWs the Stokes shift can be very easily overestimated if PL spectra are compared with ER spectra since the intensity of fundamental transition observed in ER spectra significantly decreases with the increase in QW width. In this way ER signal related to excited states can dominate in ER spectra. This effect is strongly enhanced by QW inhomogeneities.

# 1. Introduction

Polar InGaN quantum wells (QWs) are industry important structures because of their applications in blue-green light emitting diodes (LEDs) and laser diodes (LDs).[1–6] Due to polarization effects in this QW system[7,8] a strong built-in electric field is present in polar InGaN QWs that decreases the overlap of electron and hole wavefunctions with increasing QW width. Therefore, commonly very thin QWs ($d < 3$nm) are used for LED and LD applications. In this regime of QW thicknesses the width fluctuation of even a few monolayers (MLs) significantly affects energies of QW transitions and enhances the broadening of QW emission.[9] In addition, fluctuations of indium concentration also affect energies of QW transitions. These phenomena are especially important for green InGaN LEDs and LDs because of higher indium concentrations, i.e., larger strains and hence stronger piezoelectric fields incorporated in these devices.

Absorption-like experiments such as electromodulated reflectance (ER) spectroscopy (i.e., photoreflectance, electroreflectance or contactless electroreflectance) together with photoluminescence (PL), which is an emission-like method, are applied to study the optical quality of QWs,[10–12] are used in this study. PL is sensitive to localized states, which are manifested mainly at low temperatures, and, therefore, the temperature-dependent PL measurements are performed in order to evaluate the scale of carrier localization in the investigated samples.[13–21] The carrier localization can be also evaluated by the comparison of PL with absorption-like spectra.[12,22,23] ER spectroscopy is a powerful technique to study energies of QW transitions.[24,25] For non-polar QWs the comparison of PL with ER spectra is rather simple to interpret[24] due to the selection rules for transitions observed in absorption-like spectra. For polar QWs such a comparison is quite a complex issue because there are many more transitions between excited states that contribute to the ER spectrum. The signal is then composed of many closely spaced resonances, which overlap with the signal related to the fundamental transition and with each other. This complexity is further enhanced for inhomogeneous QWs.

Up to now ER spectroscopy was applied a few times to study InGaN QWs[26–31] but the problem of superposition of QW transitions related to different states present in ER spectra and the broadening of QW transitions associated with QW inhomogeneities was not carefully addressed. In general, such an issue is problematic to study experimentally since it is difficult to control independently QW fluctuations, i.e., to have a set of samples with fluctuating QW width without QW content fluctuations, and conversely to have a set of QWs with fluctuating

QW content and without QW width fluctuations. However, this issue can be studied carefully theoretically within the random quantum well (RQW) model, which has been proposed in Ref. 32 to calculate the broadening of interband and intersubband transitions in GaN/AlN QWs.

In this work the RQW model has been applied to analyze the broadening of interband transitions and the Stokes shift in InGaN/GaN QWs dedicated for green/blue lasers. This study aims to describe the impact of inhomogeneous broadening (IHB) on PL and ER spectra of polar and non-polar InGaN QWs, and to establish guidelines regarding the proper interpretation of measured spectra. The paper is structured as follows. In the next section we describe the theoretical model for simulating the spectra of InGaN QWs with structural inhomogeneities (QW width and content fluctuations). Simulated PL and ER spectra for polar and non-polar QWs are analyzed in Section 3, focusing on the effects of (a) increasing inhomogeneities magnitude, (b) nominal well thickness, and (c) nominal well alloy composition. Section 3 (d) discusses the influence of phase factor on ER spectra. Section 3 (e) is devoted to interpret experimental spectra by comparing them with simulated ones. Section 4 summarizes our findings.

## 2. Theoretical model

We study inhomogeneous QW structures consisting of an $In_xGa_{1-x}N$ active layer of varying thickness $d$ and indium content $x$, and a GaN barrier with thickness $b = 10$ nm. The inhomogeneities (illustrated in Fig. 1) are modeled within the RQW theoretical framework[32,33] by an ensemble of $N$ homogeneous structures with thickness and composition taken randomly from given distributions.

For each homogeneous structure, the band edge profiles of the valence band (VB($z$)) and conduction band (CB($z$)) are calculated. Some details of the calculation method can be found in previous works.[33,34] The growth direction $z$ is along the $c$ wurtzite crystal axis. Pseudomorphic growth on a GaN substrate is assumed for strain calculation. The built-in electric field in a multiple QW (MQW) is calculated with periodic boundary conditions. We consider undoped QWs and neglect the screening of electric field caused by residual carrier concentration. For simplicity, we consider only heavy hole levels since their separation from other hole levels is small compared to the broadening energy.

The non-polar wells are simply obtained by equating the total polarization of every layer to zero.[35] Although real structures grown along non-polar directions $a$ or $m$ would differ slightly due to the anisotropy of material parameters, this approach facilitates direct comparison with polar QW since energy gaps and effective masses remain unchanged.

All parameters used in electronic structure calculations are collected in Table I. We use the parameters from Ref. 36 and the updated recommended values for deformation potentials, piezoelectric constants, and InN bandgap from Ref. 37. Valence band maximum offset for unstrained InN/GaN interface (natural VBO) is 0.7 eV with linear interpolation[38], which corresponds to band discontinuity ratio $\Delta VB/\Delta E_g \approx 0.18$, where $\Delta E_g$ is the difference between bandgaps. The InGaN bandgap bowing is 1.7eV.[39]

The Shrödinger equation is solved using tridiagonal matrix method to find energies $E^e(i), E^h(j)$ and envelope wavefunctions $e_i(z), h_j(z)$ of at most 2 electron and 3 hole lowest confined states: $1 \leqslant i \leqslant 2$, $1 \leqslant j \leqslant 3$. Telling if a state is confined in a polar MQW may be ambiguous, since the top of a polar QW is not strictly defined. We propose a criterion that relies on the comparison of energy level with band discontinuity at the central point $z = z_{1/2}$ of the well layer. Namely, an electron or hole state is treated as confined if $E^e(i) < CB\left(z_{1/2}\right) + \Delta CB$ or $E^h(j) > VB\left(z_{1/2}\right) - \Delta VB$, respectively. In other words, it is confined by at least half of the barrier width. This method is consistent with non-polar QWs and leads to transition energy between confined states being always lower than the barrier bandgap.

Finally, the resulting set of the calculated transition energies $E_{ij} = E^e(i) - E^h(j)$ and transition oscillator strengths $f_{ij} = \left|\int h_j^*(z)\, e_i(z)\, dz\right|^2$, serves to calculate the PL and $\Delta R/R$ spectra

$$\text{PL}(E) \propto \sum_{i,j} \frac{f_{ij}\, \Gamma}{(E-E_{ij})^2 + \Gamma^2}, \qquad (1)$$

$$\frac{\Delta R}{R}(E) \propto \text{Re} \sum_{i,j} \frac{f_{ij}\, \Gamma\, e^{i\theta_{ij}}}{(E-E_{ij}+i\Gamma)^m}, \qquad (2)$$

where we use the homogeneous broadening parameter $\Gamma \approx k_B T \approx 25$ meV, and $m = 2$ relevant to excitonic transitions.[40] We neglect other factors that influence the intensity of PL, like excitation efficiency and non-radiative decay rate. Parameters $\theta_{ij}$ denote phase of transition resonances. It is related (among others) to the ratio of the distance between the QW and the sample surface to the wavelength and is difficult to calculate.[40–42] To simplify the analysis, we take $\theta_{ij} = \theta = -\pi/4$ for all transitions and we discuss the influence of $\theta$ on the modulation reflectance spectra in Section 3 (d).

The final PL or ER spectrum for an inhomogeneous structure is the average of spectra for $N$ homogeneous structures, each calculated as described above. The ensemble of $N = 5 \times 10^4$ homogeneous QW structures is simulated with well thickness and concentration taken from Gaussian distributions with given expected values $d$, $x$, and standard deviations $\Delta d$ and $\Delta x$ respectively.

As a parameter describing quantitatively the IHB of each optical transition, we propose to calculate the individual moduli of ER resonances taking into account the inhomogeneities

$$\Delta \rho_{ij}(E) = \frac{1}{N}\sum_{k=1}^{N} \frac{f_{ij}^{(k)} \Gamma}{\left(\left(E-E_{ij}^{(k)}\right)^2 + \Gamma^2\right)^{n/2}}, \quad (3)$$

where the superscript $(k)$ refers to the quantities calculated for the $k$-th of $N$ simulated structures. From $\Delta \rho_{ij}$ we calculate the full width at half maximum (FWHM) range of energy $[E_1, E_2]$. It is defined as the smallest interval such that

$$\forall_{E \notin [E_1, E_2]} \Delta \rho_{ij}(E) < \frac{1}{2}\max \Delta \rho_{ij}. \quad (4)$$

In case of homogeneous broadening the FWHM is $E_2 - E_1 = 2\Gamma \approx 50$ meV for each transition. Since the energy of an optical transition can be more reliably calculated than its intensity, it may be more accurate to describe IHB by the FWHM parameter than by the variation of amplitude of broadened resonances.

## 3. Results and discussion

In this section we analyze the influence of inhomogeneities on the ER and PL spectra of different QWs. The results in Sections 3 (a)-(d) are based on simulations within the RQW model. In Section 3 (e) the experimental PL and contactless ER spectra measured for an inhomogeneous MQW sample are interpreted via comparison with corresponding calculated spectra.

**a) Influence of increasing inhomogeneities**

The influence of fluctuations on the optical spectra of inhomogeneous polar and non-polar QWs is presented in Fig. 2. We consider polar (PQW) and non-polar (NQW) variant of $In_xGa_{1-x}N(d)/GaN(10nm)$ QW with $d = 3$ nm, $x = 25\%$. Figure 2 (a) shows their electronic structures and tables of oscillator strengths $f_{ij}$. In PQW the fundamental transition energy $E_{11} < E_g(In_xGa_{1-x}N)$ due to strong quantum-confined Stark effect. Fig. S1 in Supplementary Material shows the case of $d = 2$ nm, $x = 15\%$, where $E_{11} > E_g(In_xGa_{1-x}N)$, which leads to similar conclusions.

Figure 2 (b) shows the simulated PL and ER spectra of homogeneous QWs with transition energies indicated. PL is centered around the fundamental transition while ER represents all allowed transitions. The key differences between the ER spectra of homogeneous polar and non-polar QWs, which determine their different evolution when the magnitude of fluctuations increases, are the following. We describe them briefly, although they are broadly known in literature.

In absence of a built-in electric field the square infinite QW approximation holds. Due to the symmetry of CB and VB profiles, the envelope wavefunctions in CB and VB wells are mutually orthonormal, and thus the transition oscillator strengths (see the table in Fig. 2 (a) for NQW) satisfy the relation $f_{ij} = \delta_{ij}$ (the Kronecker's delta). Consequently, in Fig. 2 (b) for NQW we observe only 2 resonances with comparable amplitudes, corresponding to the transitions h1–e1, h2–e2. We point out that the transitions are relatively well separated in energy: by the sum of excited state separation energies in CB and VB wells.

For polar QWs, contrarily, the profiles of VB and CB wells differ, thus the previous symmetry is broken, see Fig. 2 (a) for PQW. In particular, the potential well minima for electrons and holes lie at the opposite interfaces of the InGaN layer. The electron envelope functions are no longer orthonormal to the hole ones. This results in a multitude of allowed transitions $f_{ij} \neq 0$, see the table in Fig. 2 (a) for PQW. In principle, we can see in ER spectrum all $2 \times 3$ resonances of varying amplitude. The fundamental transition h1–e1 is weaker than some higher-energy transitions e.g. h1–e2.[43] The sum rule (see e.g. Eq. (32) in Ref. 44) indicates that there must be significant overlap between even higher states, because the sums in columns and rows of our $f_{ij}$ table are much lower than one. The overall amplitude of spectra is lower than in the case of NQW due to incomplete overlap between hole and electron states. And most importantly, the interband transitions contributing to ER spectra are weakly separated in energy. This is mainly because excited hole states are very close in energy, due to essentially triangular shape of the well (see PQW in Fig. 2 (a)), and transitions from all of them to a given electron level are allowed.

Having compared the spectra of homogeneous QWs, we proceed to describe the influence of inhomogeneities which are of major interest to this paper. Figure 2 (c)-(e) shows how the PL and ER spectra evolve when we increase the range of composition fluctuation $\Delta x$, width fluctuation $\Delta d$, and both at the same time, respectively. For all three cases of increasing IHB, the PL broadens while its peak intensity decreases, keeping the integrated PL intensity almost the same as for the corresponding homogeneous QWs (Fig. 2 (b)). The ER spectra also broaden but their amplitude decreases much more rapidly than that of PL, especially for PQW (note the magnification of some $\Delta R/R$ plots). Investigation of this effect, which is observed in experiment, is the main subject of this study. The advantage of using computer simulation for this purpose is not just an independent analysis of different sources of fluctuations, but also the ability to identify the contributions from different transitions into the ER spectrum.

In Fig. 3 the simulated ER spectra are supplemented by the broadening ranges at half-maximum of individual transitions contributing to the spectra, calculated according to Eq. (4). For any fluctuation range $\Delta x, \Delta d$ (left axis) one can read (on the bottom axis) the IHB energy intervals corresponding to each allowed interband transition. This figure illustrates the underlying mechanism behind the evolution of spectra described in Fig. 2. We can clearly see the qualitative difference between non-polar and polar QWs. In NQW, each resonance broadens independently because the energy separation between transitions is much larger than the broadening width. In PQW, contrarily, this independent broadening is possible only for very small fluctuations, since there are more transitions and energy separations are much smaller. Even small fluctuations cause the resonances to overlap and "interfere" destructively. Since each resonance consists of positive $\left(\frac{\Delta R}{R} > 0\right)$ and negative $\left(\frac{\Delta R}{R} < 0\right)$ parts, the negative part of signal from one resonance cancels the positive from the adjacent one.

As a result, the low-strength overlapping resonances are the first to disappear when even small inhomogeneity exists. For example, the close transitions h2–e1 and h3–e1 overlap for $\Delta x > 0.25\%$ In (Fig. 3 (b)) and for $\Delta d > 0.25$ ML (Fig. 3 (d)). This effect explains the evolution of ER spectra for PQW around 2.4-2.5 eV in Fig. 2 (c)-(e).

For large inhomogeneity in Fig. 3 (b), (d) we can see that all transitions overlap. Consequently, the ER resonances with higher oscillator strength melt into a broad and flat resonance-like shape between 2.0-3.0 eV, visible also in Fig. 2 (c)-(e) for PQW with the largest inhomogeneities. We point out that it is not centered around the fundamental transition (2.27 eV) but rather around h1–e2 transition (2.65 eV) which is the strongest in this QW.

Transitions in PQW are more strongly broadened than their corresponding transitions in NQW. In particular, the h1–e1 transition is 1.5 times broader in PQW than in NQW for $\Delta x = 2\%$ (Fig. 3 (a), (b)), and it is 3.2 times broader for $\Delta d = 2$ ML (Fig. 3 (c), (d)). It means that the built-in electric field amplifies the transition energy dependence on QW parameters.

One can also distinguish subtle differences between the ways how the increase of $\Delta x$ or $\Delta d$ broadens the spectra. The content fluctuation $\Delta x$ (Fig. 3 (a), (b)) leads to qualitatively similar broadening for both polar and non-polar QWs (see also Fig. 2 (c)). For QW width fluctuation $\Delta d$, contrarily, comparison of Fig. 3 (c) with (d) clearly shows quantitative and qualitative differences. Namely, width fluctuation leads to much stronger and asymmetric IHB in PQW. In particular, the h1–e1 transition in Fig. 3 (d) is broadened very asymmetrically, i.e.

the broadening toward lower energies is limited. As a result, the PL peak is blue-shifted in inhomogeneous PQWs in Fig. 2 (d) and (e). This effect results from the strong dependence of transition oscillator strengths $f_{ij}$ on the QW width $d$.

In the cases shown in Fig. 3 (a), (b) and (d) the fundamental transition is the broadest among all transitions. But interestingly, in Fig. 3 (c) the excited state transition is much more broadened than the fundamental one. Figure 2 (c) for NQW also shows that the h2–e2 transition is very sensitive to $\Delta d$ broadening. According to infinite square QW model, the energy of higher transitions varies much more with the width fluctuation: $\frac{\partial E_{ii}}{\partial d} \propto i^2$. This explains why for non-polar inhomogeneous QWs with realistic inhomogeneity like 1 ML and 1% In, the transition between excited states may be hardly visible.

**b) Constant inhomogeneity and increasing In content**

In the previous subsection we considered only one example QW structure in its polar and non-polar variant. In this and the following subsection we compare the influence of constant inhomogeneity range ($\Delta x = 1\%$ In and $\Delta d = 1$ ML) on polar QWs with varying nominal parameters. In Fig. 4 we show the effect of changing indium content $x$ in the range from 10% to 30%. Each of the three columns illustrates the case for a selected value of well width. Top panels (a)-(c) show the band profiles and the ground electron and hole wavefunctions for the cases with extreme values of $x$. The evolution of spectra with changing $x$, shown in bottom panels (d)-(f), is similar for all three considered well widths. When In content increases, the built-in electric field is stronger and the transition energy decreases. Therefore the spectra shift towards red.

For QWs with higher $x$, PL intensity is lower and it is a little more broadened. The increasing built-in electric field makes the wavefunctions more confined, therefore their overlap and PL intensity decrease, especially in wider QWs (Fig. 4 (f)). Higher built-in electric field is also responsible for the enhanced broadening.

We see that for higher $x$ the $\Delta R/R$ spectrum spans a larger interval of energy but has a lower amplitude. It seems counterintuitive that the ER spectrum is more attenuated for higher nominal indium content $x$ although the *relative* fluctuation $\Delta x/x$ is lower. This is because the higher the electric field, the more sensitive the spectrum is to IHB. In particular, deeper QWs can confine more states and allow more interband transitions. The overlapping resonances can then interfere destructively, as described in previous section, leading to the attenuation of ER spectra.

### c) Constant inhomogeneity and increasing QW width

In this subsection we consider QWs with the same inhomogeneity range ($\Delta x = 1\%$ In, $\Delta d = 1$ ML) and discuss effects of changing QW width $d$ in the range from 1.0 nm to 3.5 nm. This case is illustrated in Fig. 5 for three selected values of $x$. The PL intensity decreases strongly with increasing $d$ because the fundamental oscillator strength $f_{11}$ rapidly decreases for thick wells. The $\Delta R/R$ spectrum red-shifts and exhibits lowering of amplitude as thickness $d$ increases. Furthermore, for thicker QWs the low-energy part of the spectrum is hardly visible, because the wavefunctions of electron and hole lower levels are separated. On the other hand, the transitions from excited states become more important. As a result, the ER spectra are significantly non-zero in a similar range of energy 2.5-3.3 eV, only their shape change and the amplitude of $\Delta R/R$ decreases for larger $d$. This effect is not caused by increasing electric field, which does not change significantly with $d$, but rather by the decrease of oscillator strengths with increasing well width, like in the case of homogeneous QWs.

### d) Influence of the phase of ER resonances

It is difficult to select the value of the phase parameter $\theta$ in Eq. (2) that should be used to simulate the modulation reflectance spectra.[45] The effect of phase parameter on the ER spectra is presented in Fig. 6. Firstly, the phase modifies the shape of resonances. For $\theta = 0$ (respectively, $\theta = -\pi/2$) the spectrum consists of resonance peaks with even (odd) symmetry. For $\theta = -\pi/4$ and in general case there is no such symmetry. From Eq. (2) it follows that the areas above and below the level $\Delta R/R = 0$ are always equal, although this property is not clearly visible on the plots. Secondly, the change of $\theta$ apparently shifts the energy region of the resonance: the larger the broadening, the larger the shift. One must be aware of this effect when reading the transition energy from the position of a resonance peak in a spectrum, and that it is generally difficult to obtain accurate transition energy without proper fitting. But what is important for the present analysis is that the value of $\theta$ does not qualitatively change the evolution of spectra with increasing IHB. In particular, the ability to resolve different peaks in spectra is similar regardless of $\theta$. For instance, in Fig. 6 (b) the ability to distinguish resonances h2–e1 and h3–e1 is comparable for different $\theta$. The same applies to h2–e2 and h3–e2.

A complication for generating modulation reflectance spectra arises because the phase is unknown a priori and may be different for each transition in the spectrum. Moreover, its dependence may also change with QW width $d$ and composition $x$.

Another issue is a possible fluctuation of $\theta$ itself resulting from e.g. inhomogeneity of overlayer thickness and refractive index. This effect can be shown to be negligible in comparison to the IHB caused by typical $\Delta d$ and $\Delta x$ since the characteristic dimensions of QW structure are much shorter than the wavelength.

Since our aim is not to reproduce ER spectra of particular QWs but rather to simulate their IHB, we decided to use a constant value $\theta = -\pi/4$ for all transitions to simplify the analysis.

**e) Comparison with experiment**

The above theoretical analysis allowed us to separately analyze different broadening-causing factors. In this section we demonstrate that our theoretical predictions are consistent with the experiment. To this end, we compare the experimental and generated spectra of polar QWs with inhomogeneities.

Figure 7 shows a scheme of the investigated sample. It was grown using plasma-assisted molecular beam epitaxy under In excess with Ga flux 1.45 um/h and N flux of about 2.6 um/h, on Ga-polar ammonothermal GaN substrate with 300 nm epitaxial GaN buffer layer. The heterostructure consists of three InGaN QWs of 3 nm nominal width and indium concentration of about 24%. The QWs are separated by 8 nm thick $In_{0.08}Ga_{0.92}N$ layers. The MQW structure is sandwiched between another 30 nm thick $In_{0.08}Ga_{0.92}N$ layers to separate the structure from the surface and GaN/InGaN interface.

Photoluminescence measurement was carried out at room temperature using HeCd laser operating at 325 nm. For contactless electroreflectance (CER) measurements the electric field in the sample is modulated by a capacitor with one semi-transparent copper-wire mesh electrode with applied alternating voltage (at ~280 Hz). White light from a halogen lamp is reflected from the sample surface through a monochromator to a photomultiplier. A lock-in technique is used for measurements of relative reflectance changes $\Delta R/R$. More details can be found in a previous report.[31]

Results of room temperature CER and PL measurements are presented in Fig. 8 (a), (e). They are compared with simulations for an $In_{0.24}Ga_{0.76}N(3nm)/In_{0.08}Ga_{0.92}N(8nm)$ QW structure (which is similar to the previously discussed case of PQW from Fig. 2). The range of structural inhomogeneities is taken as $\Delta d = 1$ ML, $\Delta x = 1\%$ In because the resulting width of simulated PL is consistent with measured results. The experimental PL spectrum is very well reproduced by the simulated one.

For barriers with 8% In, assuming that they are unstrained, $E_g^{\text{barrier}}(300K) \approx 3.09$ eV. This is shown as a vertical line in Fig. 8 (a). A Franz-Keldysh oscillation above $E_g^{\text{barrier}}$ is clearly visible indicating a presence of a strong built-in electric field in the barrier. But the most important observation is that below $E_g^{\text{barrier}}$ the transitions from the polar MQW are hardly visible. It is because the energies of different transitions in the ensemble strongly overlap, see Fig. 8 (d). This in turn causes strong attenuation of QW modulation reflectance signal as shown in Fig. 8 (c). For many inhomogeneous polar QW samples, the broadened ER spectrum amplitude is lower than the noise.

For this sample, however, IHB is sufficiently low that a weak CER signal around 2.7 eV remains. From Fig. 8 (b)-(d) we see that ripples occur at the same energy interval as in the generated spectrum. It is clear that this ripple is not related to the fundamental transition located around 2.45 eV. Using transition strength histograms (Fig. 8 (d)) one can try to assign which transitions contributed to the ripple. What is observed in experimental CER around 2.75 eV is probably a trace of h1–e2 transition that would be the strongest in a homogeneous structure.

We point out that the spectra in Fig. 8 (b) differ by shape because the transitions in experimental CER spectrum apparently have different phase angles than the value assumed in simulation – this issue was discussed in Section 3 (d).

The absence of sharp well-resolved QW-related interband transitions in measured ER spectra of polar QWs has therefore been explained by strong broadening and attenuation resulting from structural inhomogeneities.

It is worth noting that very similar CER spectra have been observed for other InGaN QWs with similar QW width and indium concentration. Regarding experimental data it is rather impossible to have a set of QW samples with negligible QW width fluctuations and different fluctuations of indium concentration and vice versa. Therefore the comparison of theoretical simulations with experimental data is limited to one example.

## 4. Summary

In this work we have compared IHB in polar and non-polar InGaN/GaN QWs. We have separately analyzed different broadening-causing factors using computer simulation, and compared the model with experimental results. We have shown that IHB severely affects the spectra of polar QWs due to the following mechanisms:

(1) Polarization-induced breaking of square-well selection rules allows many interband transitions that are close in energy.

(2) Built-in electric field enhances the variation of transition energy with structural properties.

(3) Broadened transitions strongly overlap leading to "destructive interference" of neighboring resonances and strong attenuation of modulation reflectance spectra.

(4) Spectra of QWs with higher indium concentration are more sensitive to, say, 1% In fluctuation, counterintuitively, due to a higher electric field in those QWs.

(5) Fluctuations of structural parameters influence also the electron–hole overlaps, causing asymmetric broadening that leads to spectral shifts.

(6) For large broadening, the transition energy obtained from a fitting procedure to an ER spectrum relates to the transition that is just the least attenuated – it is not known a priori that this is the fundamental transition.

We identify three reasons, why for inhomogeneous PQWs the fundamental transition may be hardly detectable in ER: (1) initially lower amplitude than in NQW, (2) more broadened especially if width fluctuation is present, (3) partially overlapped and cancelled by adjacent resonances. Forgetting about these effects may lead to inaccurate determination of Stokes shift.

To conclude, in the case of non-polar QWs a relatively weak IHB acts on separate, selection-rules-allowed transitions, weakly affecting the ER spectra. For polar QWs the attenuation of $\Delta R/R$ spectra is much stronger due to a destructive mixing of many neighboring resonances. As fluctuations increase, they sequentially melt together and eventually form a very broad and flat resonance-like shape which is not necessarily centered at the fundamental transition energy (is blue shifted), and thus single-Aspnes-shape fit parameters would be difficult to interpret.

**Supplemental material**

See Supplementary Material for the analogs of Fig. 2 and 3 for the case of a thinner QW with lower indium content $In_{0.25}Ga_{0.75}N(2nm)/GaN(10nm)$. There is also a plot of FWHM for all transitions shown in Fig. 3.

**Acknowledgments**

This work was performed within the grant of the Polish National Science Centre (no. 2016/21/B/ST7/01274), the Polish National Centre for Research and Development grant LIDER/29/478/0185/L-7/15/NCBR/2016 and POIR.04.04.00-00-210C/16-00 (Team Tech) project of the Foundation for Polish Science cofinanced by the European Union under the European Regional Development Fund.

Table I. Parameters used in effective-mass band structure simulations

| Parameter | (unit) | GaN | InN | Bowing |
|---|---|---|---|---|
| $E_g$ (T=0) | (eV) | 3.510 [a] | 0.69 [b] | 1.7 [c] |
| $a_{Varshni}$ | (meV/K) | 0.914 [b] | 0.414 [b] | |
| $b_{Varshni}$ | (K) | 825 [b] | 154 [b] | |
| a | (Å) | 3.189 [a] | 3.545 [a] | |
| c | (Å) | 5.185 [a] | 5.703 [a] | |
| $m_e$ | ($m_0$) | 0.20 [a] | 0.07 [a] | |
| $m_h$ (A band) | ($m_0$) | 1.89 [a] | 1.56 [a] | |
| $a_{ct}$ | (eV) | −8.2 [b] | −2.5 [b] | |
| $a_{cz}$ | (eV) | −10.7 [b] | −7.8 [b] | |
| $D_1$ | (eV) | −3.6 [b] | −3.6 [b] | |
| $D_2$ | (eV) | 1.7 [b] | 1.7 [b] | |
| $D_3$ | (eV) | 5.2 [b] | 5.2 [b] | |
| $D_4$ | (eV) | −2.7 [b] | −2.7 [b] | |
| $c_{31}$ | (GPa) | 106 [a] | 92 [a] | |
| $c_{33}$ | (GPa) | 398 [a] | 224 [a] | |
| $e_{31}$ | (C/m$^2$) | −0.334 [b] | −0.484 [b] | |
| $e_{33}$ | (C/m$^2$) | 0.544 [b] | 1.058 [b] | |
| $P^S$ | (C/m$^2$) | −0.034 [a] | −0.042 [a] | −0.037 [a] |
| ε | ($\varepsilon_0$) | 9 | 15 | |
| VBO | (eV) | 0 | 0.7 [d] | |

[a] Ref. 36

[b] Ref. 37

[c] Ref. 39

[d] Ref. 38

**Figure captions**

Figure 1. Illustration of the two mechanisms of IHB and their influence on the transition energy $E$ and electron–hole overlap $\gamma$ of non-polar and polar quantum wells.

Figure 2. (a) Band diagrams and transition oscillator strengths $f_{ij}$ for the non-polar and polar QWs (labeled NQW and PQW, respectively). A comparison of their simulated PL and ER spectra for (b) homogeneous structures, and for varying range of (c) QW width fluctuation, (d) indium content fluctuation, and (e) both QW width and composition fluctuations. The spectra for PQW are magnified as indicated in red above them. The plots of $\Delta R/R$ for larger inhomogeneities in PQW are additionally magnified as denoted on their right hand side.

Figure 3. Inhomogeneous broadening of individual transitions contributing to the spectra from Fig. 2. Top panels (a)-(b) are for indium content fluctuation in the QW (see Fig. 2 (c)) and bottom panels (c)-(d) for QW width fluctuation (see Fig. 2 (d)). Simulated FWHM energy intervals (Eq. (4)), plotted in function of the fluctuation strength (left axis), are shown as the intersection of filled shapes with horizontal axes. The corresponding simulated ER spectra (right axis) are plotted for the homogeneous structure (top line) and selected fluctuation strengths: $\Delta x = 1, 2\%$ In and $\Delta d = 1, 2$ ML (middle and bottom lines). To facilitate visual comparison, horizontal energy axes span 0.9 eV in all panels.

Figure 4. Bottom panels: effect of changing In content $x$ between 5% and 30% on the simulated spectra of PQWs with constant inhomogeneity (1% In + 1 ML). Three values of well width are considered. All graphs are in the same scale. Top panels: band diagrams with ground-state wavefunctions for the corresponding homogeneous PQWs.

Figure 5. Bottom panels: effect of changing well width from 1 nm to 3.5 nm on the simulated spectra of PQWs with constant inhomogeneity (1% In + 1 ML). Three values of In concentration are considered. All graphs are in the same scale. Top panels: band diagrams with ground-state wavefunctions for the corresponding homogeneous PQWs.

Figure 6. Influence of the phase parameter $\theta$ (Eq. (2)) on simulated ER spectra for three cases: (a) homogeneous $In_{0.25}Ga_{0.75}N(3nm)/GaN(10nm)$ structure, (b) with small and (c) large IHB.

Figure 7. Scheme of the sample MQW structure grown using plasma-assisted molecular beam epitaxy.

Figure 8. Comparison of (left axes) experimental spectra for InGaN MQW and (right axes) corresponding ones generated for $In_{0.24}Ga_{0.76}N$(3 nm)/$In_{0.08}Ga_{0.92}N$(8 nm) with inhomogeneity $\Delta d = 1$ ML, $\Delta x = 1\%$ In. (a) Measured CER spectrum. (b) Close view of the experimental spectrum of MQW (below the barrier gap energy) and a possible corresponding simulated spectrum accounting for IHB. (c) Magnitude comparison of the plot from (b) with the idealized case of homogeneous sample. (d) Simulated moduli of different interband transitions (see Eq. (3)) labeled h-e using indices of hole and electron levels. (e) Room temperature PL spectra: experimental and simulated.

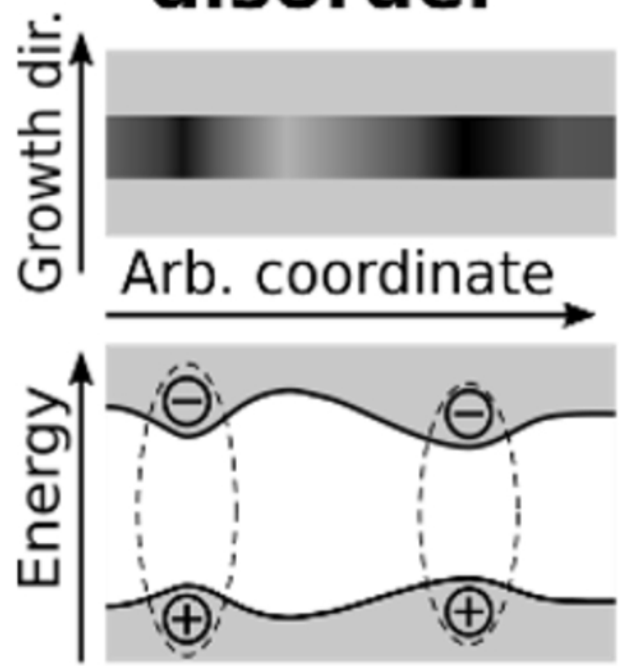
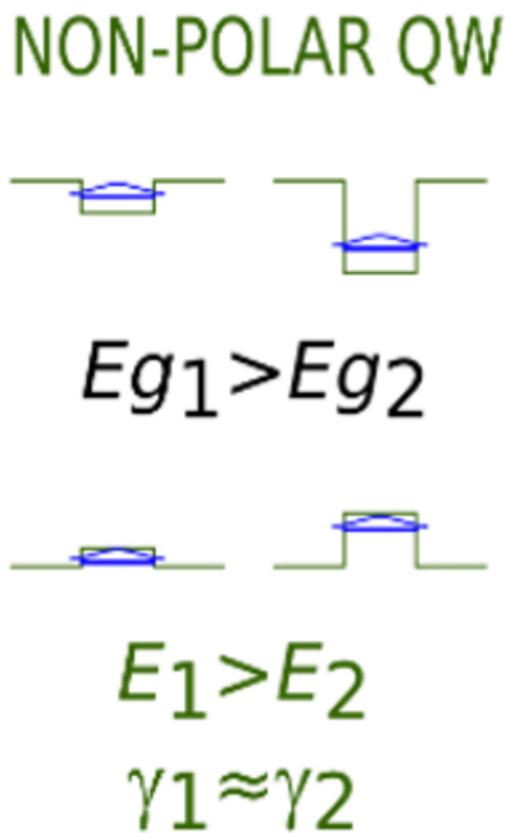
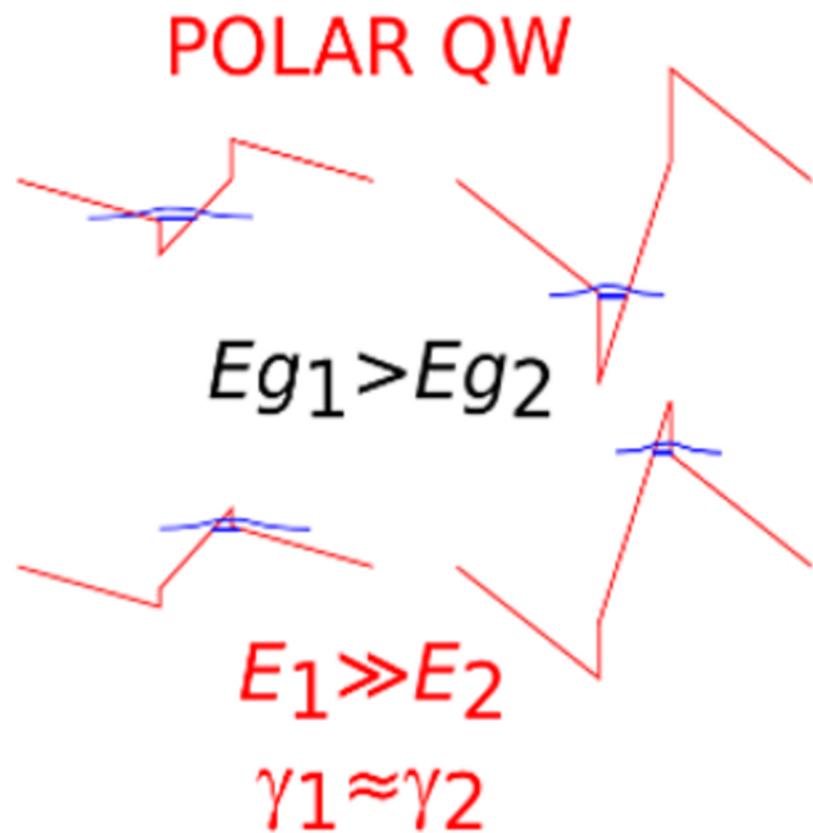
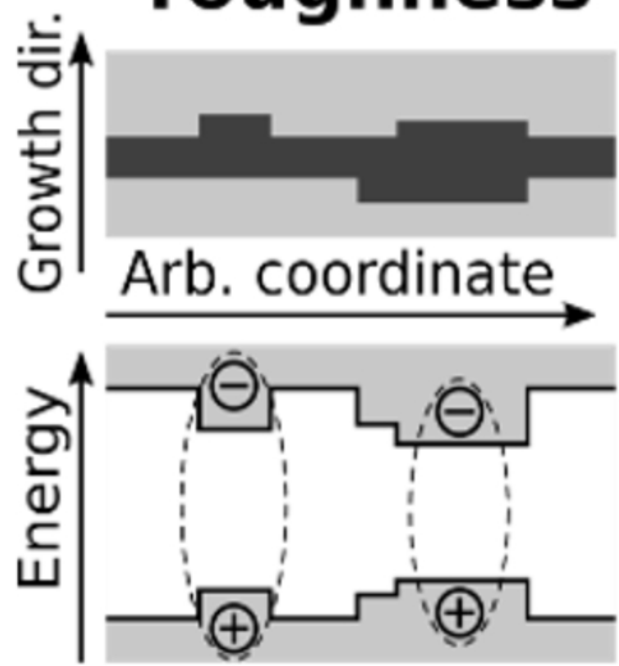
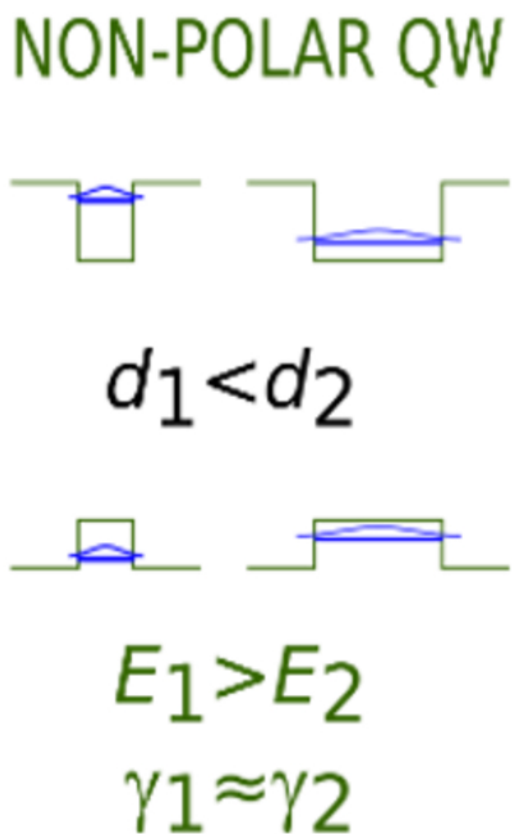
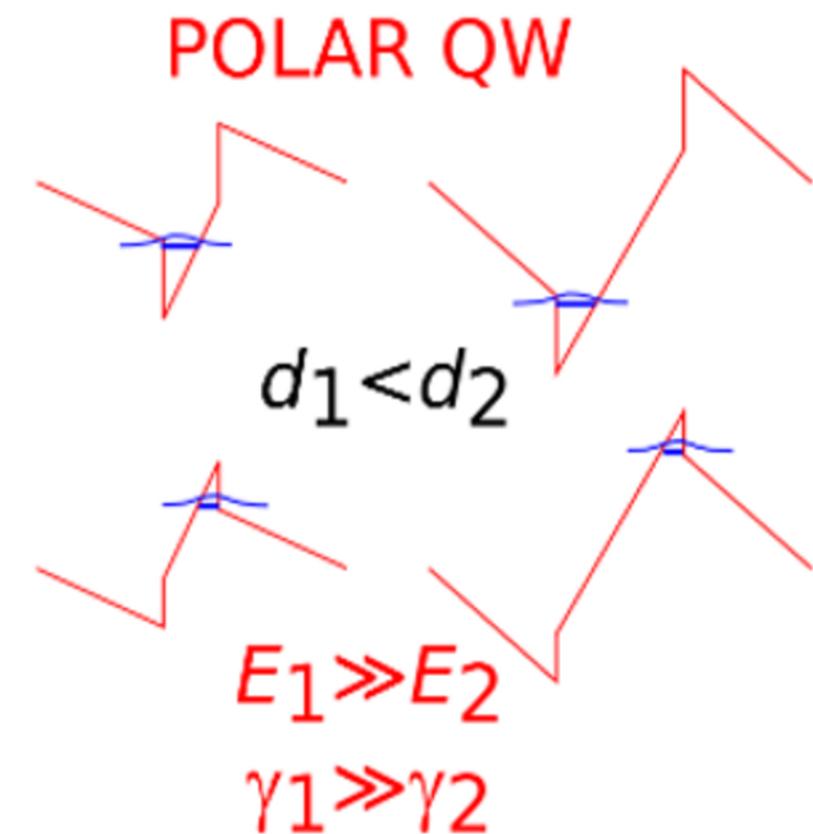

InGaN / GaN, 3 nm / 10 nm, $x = 25\%$

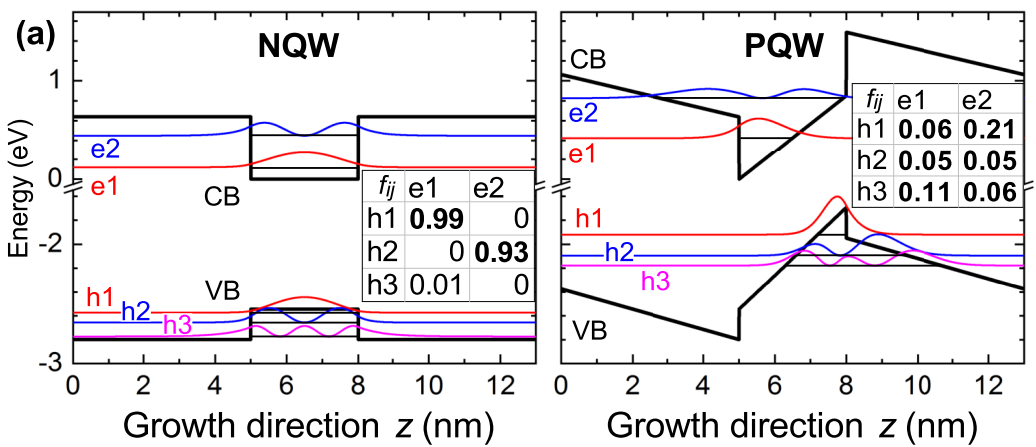

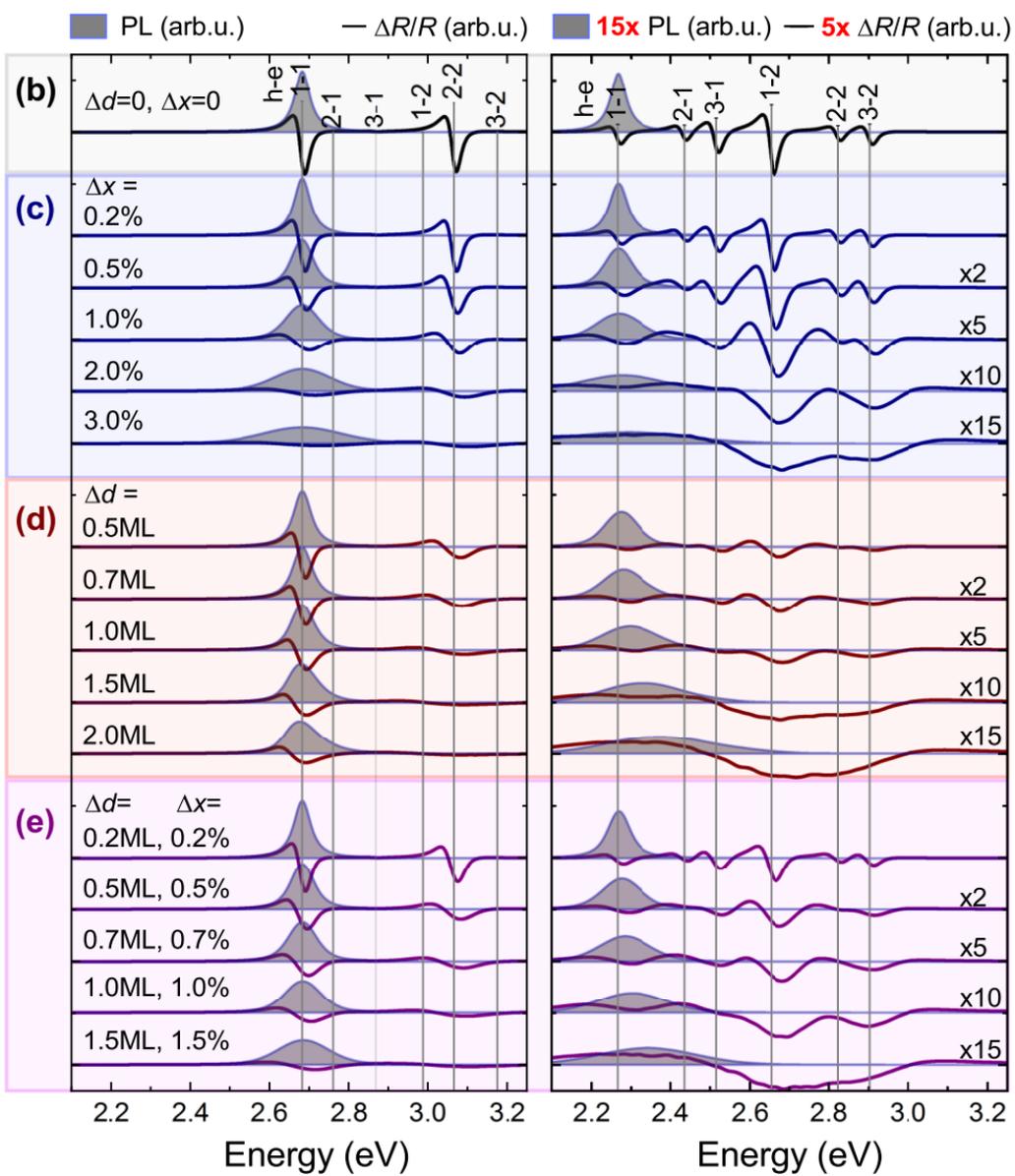

InGaN/GaN, 3nm/10nm, *x*=25%

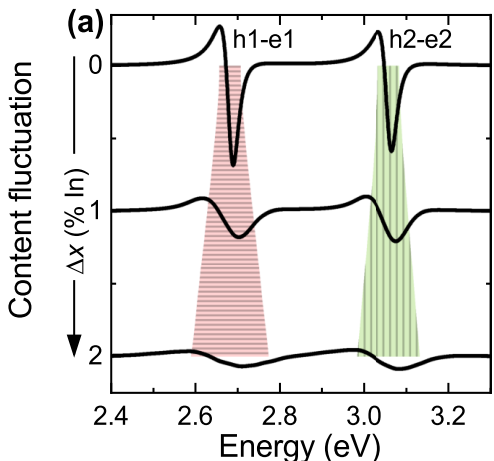
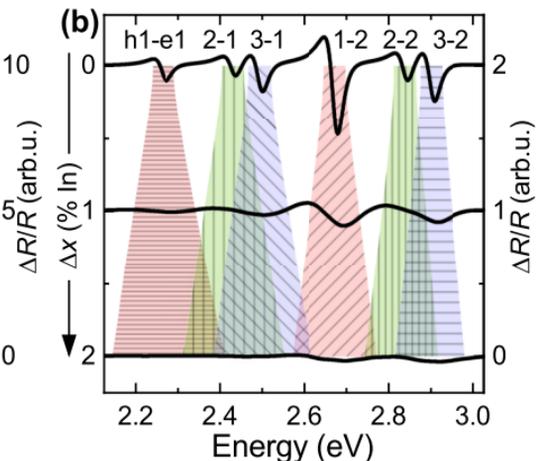
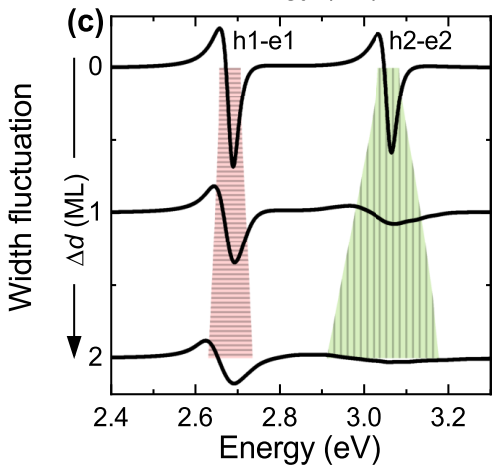
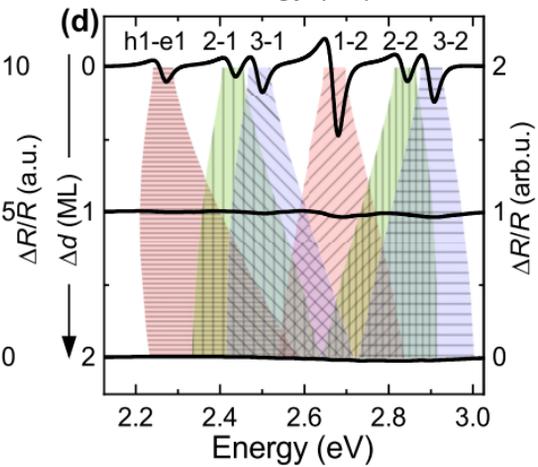

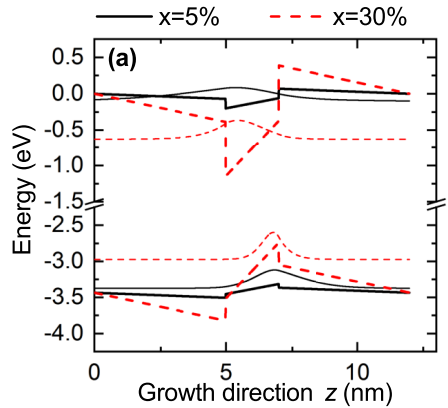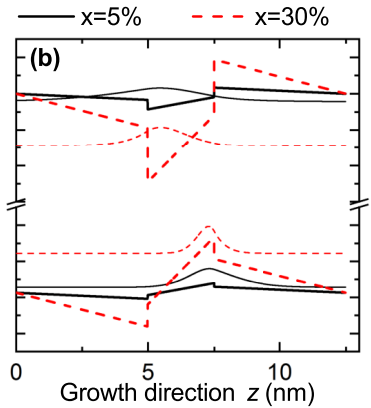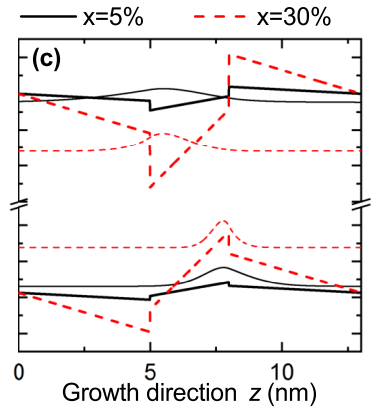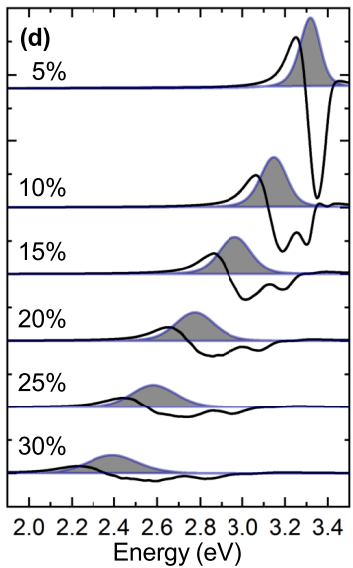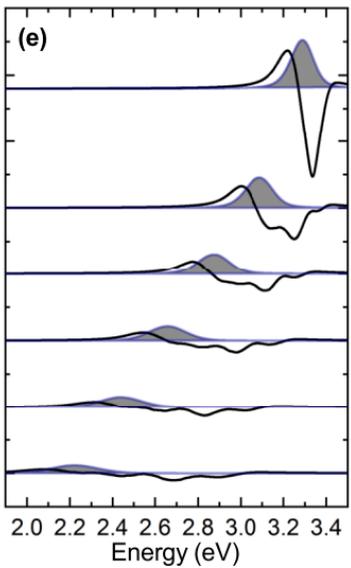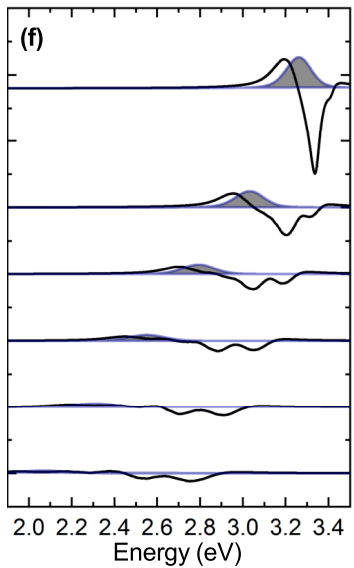

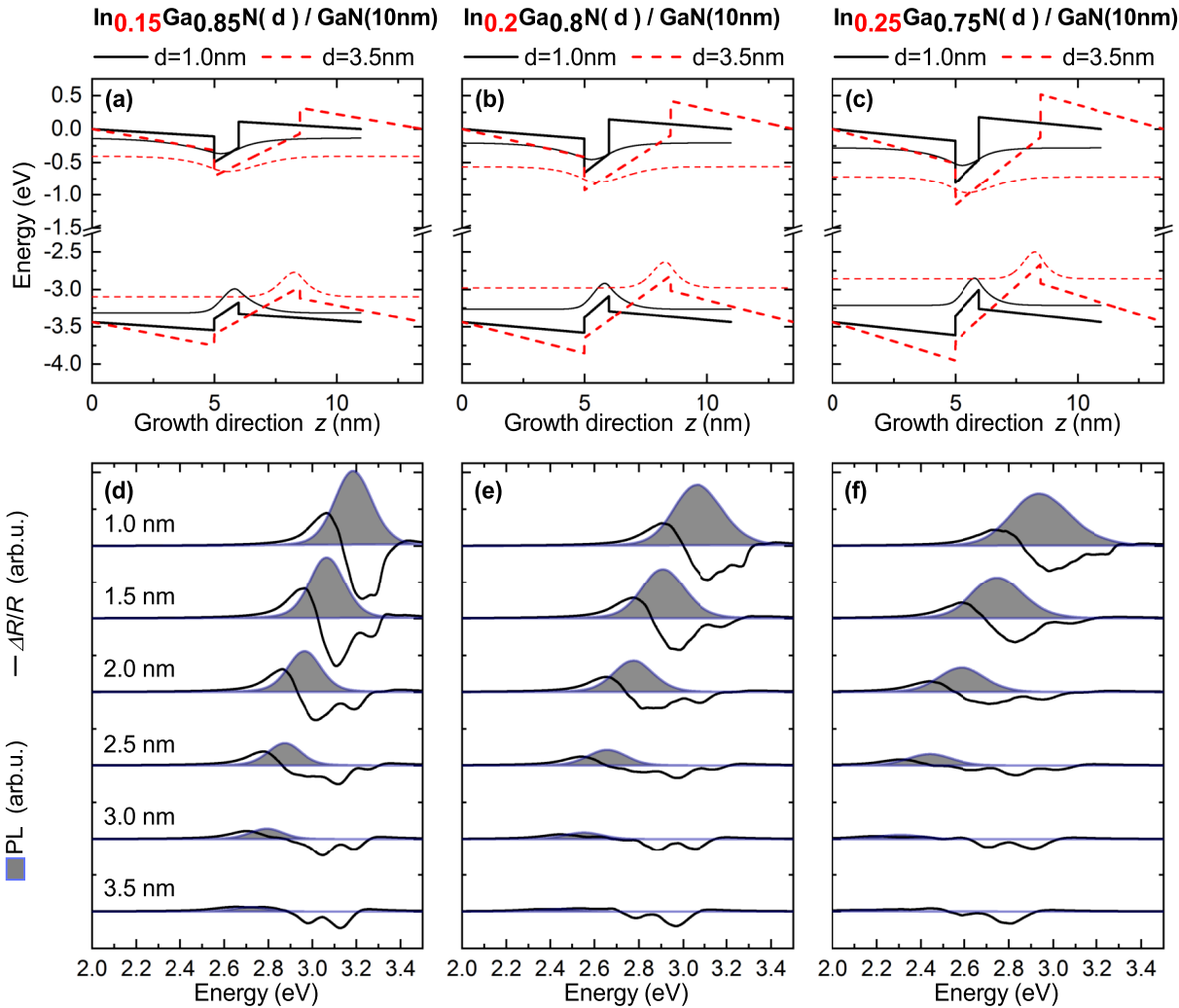

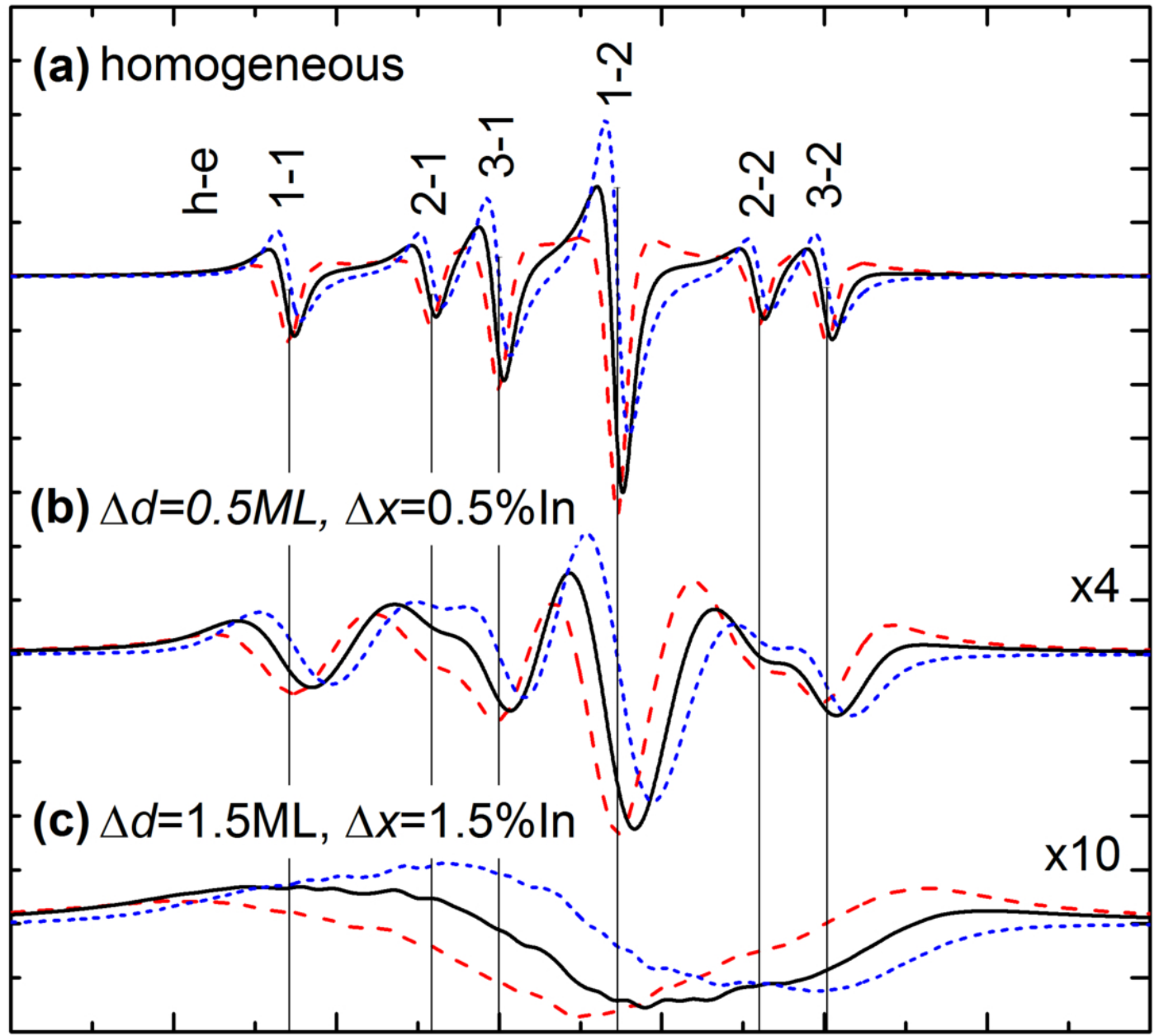

## R0346 3QW

| 22 nm InGaN ~8% | |
|---|---|
| 3 × | 8 nm InGaN ~8% |
| | 3 nm InGaN ~24% (QW) |
| 30 nm InGaN ~8% | |
| 300 nm GaN | |
| Substrate Ga-polar GaN | |

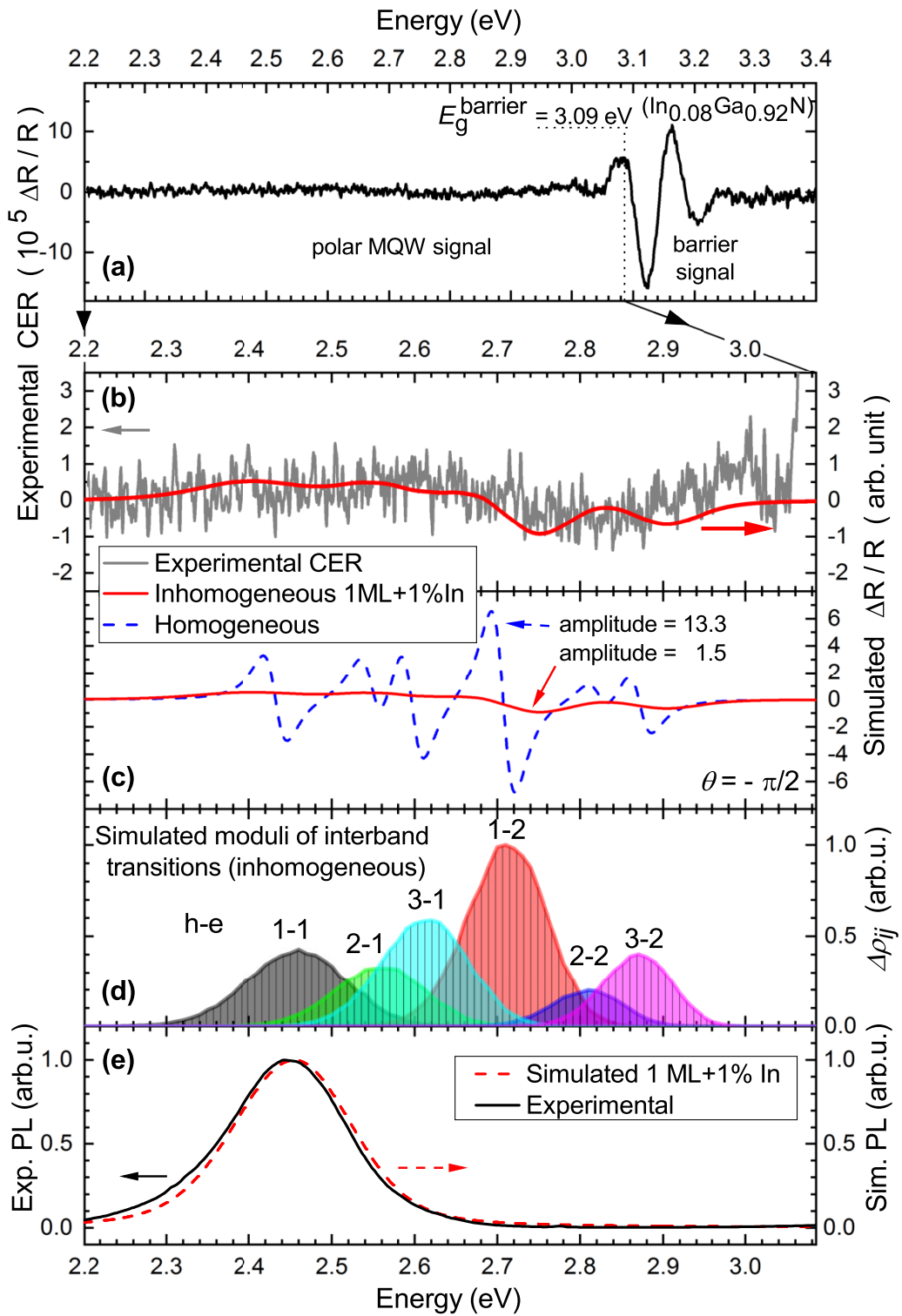

**SUPPLEMENTARY MATERIALS**

# Inhomogeneous broadening of optical transitions observed in photoluminescence and modulated reflectance of polar and non-polar InGaN quantum wells

Michał Jarema,[1] Marta Gładysiewicz,[1] Łukasz Janicki,[1] Ewelina Zdanowicz,[1] Henryk Turski,[2] Grzegorz Muzioł,[2] Czesław Skierbiszewski,[2] and Robert Kudrawiec[1]

[1] *Faculty of Fundamental Problems of Technology, Wrocław University of Science and Technology, Wybrzeże Wyspiańskiego 27, 50-370 Wrocław, Poland*

[2] *Institute of High Pressure Physics, Polish Academy of Sciences, Sokołowska 29/37, 01-142 Warsaw, Poland*

Here we present simulation results concerning the influence of increasing inhomogeneities for a different pair of non-polar and polar $In_xGa_{1-x}N(d)$/GaN(10nm) quantum wells with $d = 2$ nm, $x = 15\%$.

Figures S1 and S2 are analogous to Fig. 2 and 3 from the manuscript, respectively. Figure S3 is intended to facilitate quantitative analysis of Fig. S2 and Fig. 3 from the manuscript. Additionally, Fig. S3 (c), (f) shows that barrier width fluctuation $\Delta b$ causes only very small broadening (for the barrier width of 10 nm) which was neglected in the rest of this paper.

We conclude that the discussion from Section 3 (a) applies to this pair of QWs as well. The polarization-induced broadening enhancement is just less pronounced for these thinner QWs with lower indium content.

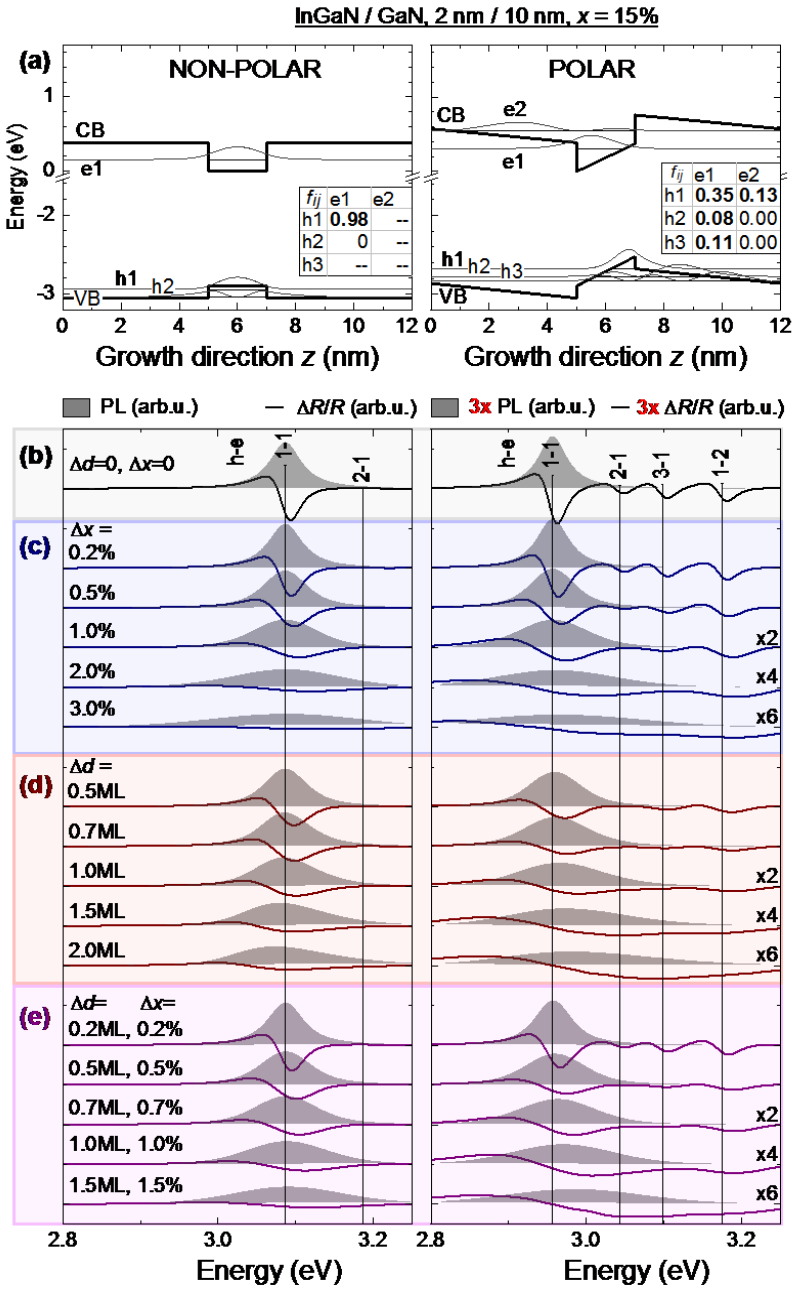

Figure S1. (a) Band diagrams and transition oscillator strengths $f_{ij}$ for the non-polar and polar QWs (labeled NQW and PQW, respectively). A comparison of their simulated PL and ER spectra for (b) homogeneous structures, and for varying range of (c) QW width fluctuation, (d) indium content fluctuation, and (e) both QW width and composition fluctuations. The spectra for PQW are magnified as indicated in red above them. The plots of $\Delta R/R$ for larger inhomogeneities in PQW are additionally magnified as denoted on their right hand side.

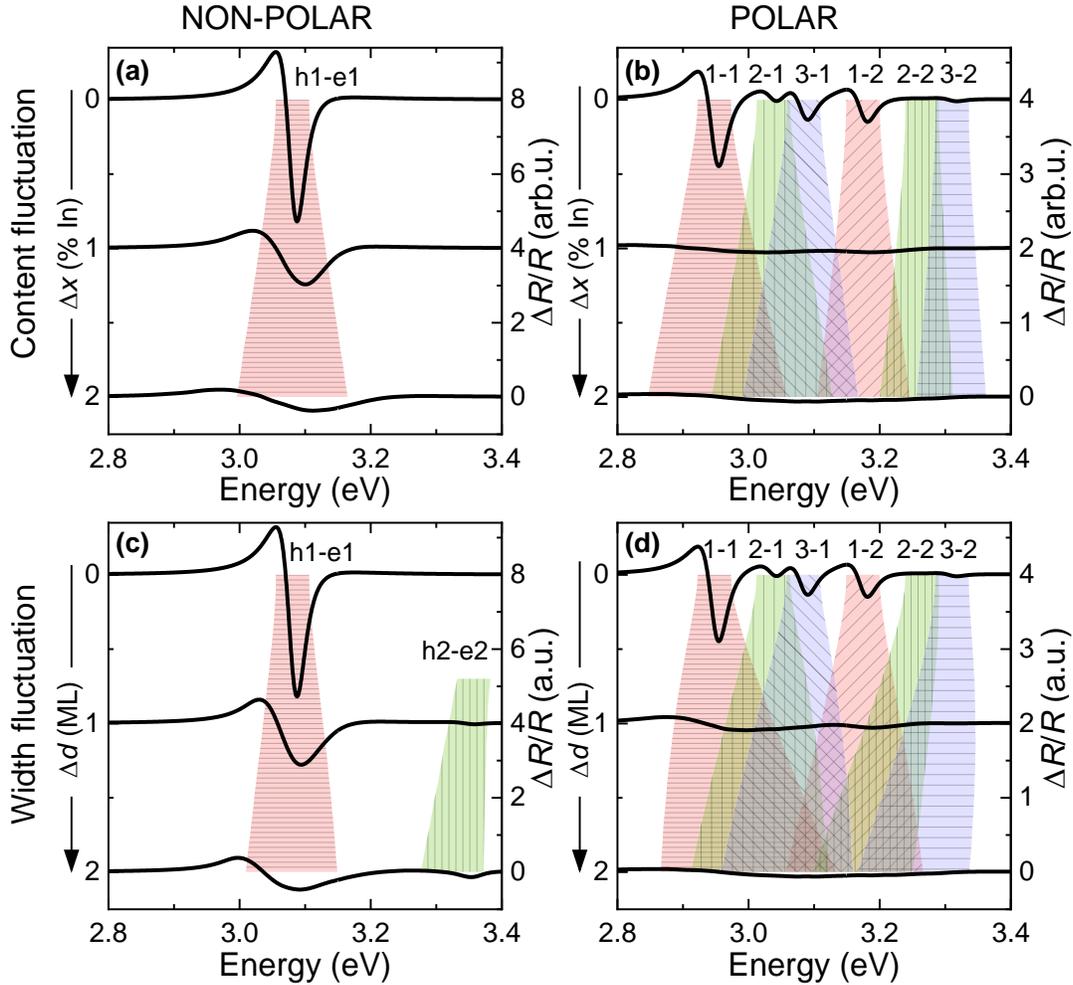

Figure S2. Inhomogeneous broadening of individual transitions contributing to the spectra from Fig. S1. Top panels (a)-(b) are for indium content fluctuation in the QW (see Fig. S1 (c)) and bottom panels (c)-(d) for QW width fluctuation (see Fig. S1 (d)). Simulated FWHM energy intervals (Eq. (4)), plotted in function of the fluctuation strength (left axis), are shown as the intersection of filled shapes with horizontal axes. The corresponding simulated ER spectra (right axis) are plotted for the homogeneous structure (top line) and selected fluctuation strengths: $\Delta x = 1, 2\%$In and $\Delta d = 1, 2$ ML (middle and bottom lines). In (c) the transition h2–e2 appears because e2 level begins to be confined in QWs when width fluctuation is strongly positive. Horizontal energy axes span 0.6 eV which is less than 0.9 eV in the analogous Fig. 3.

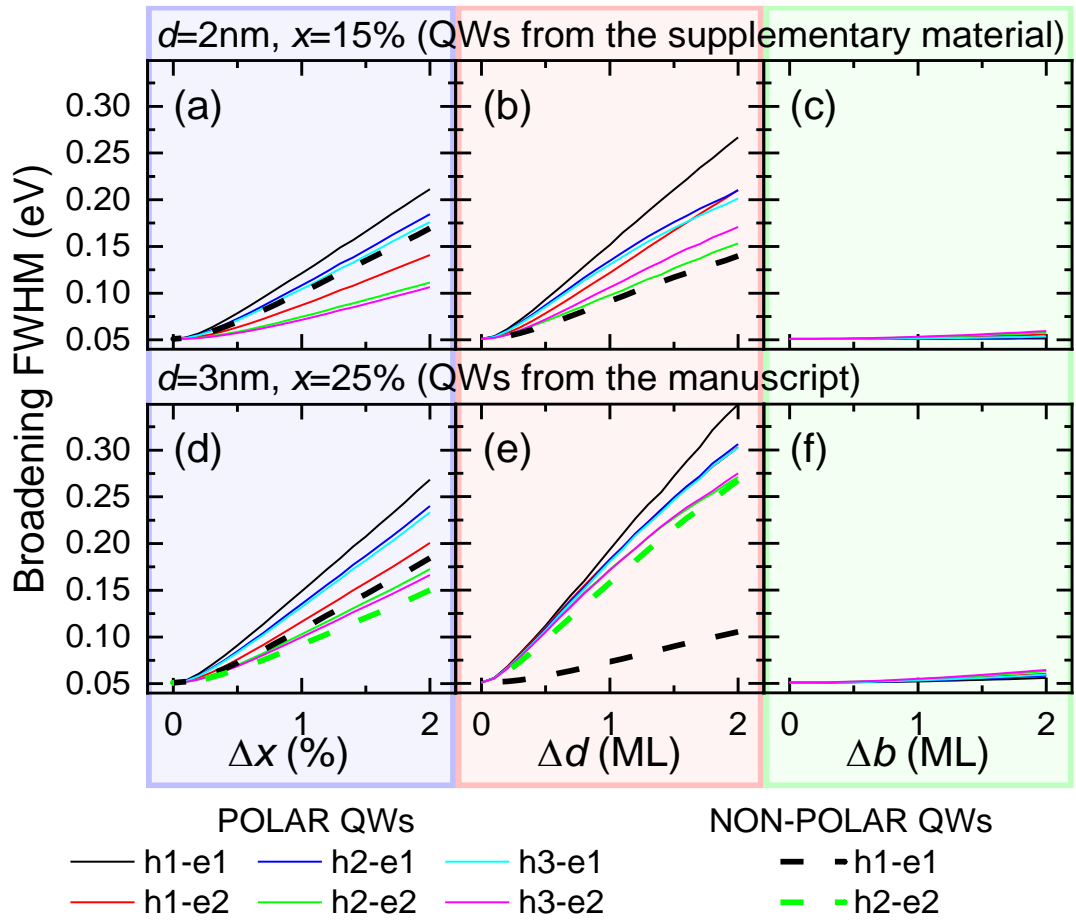

Figure S3. The broadening FWHM depending on the magnitude of fluctuation of (a, d) indium content $\Delta x$, (b, e) well width $\Delta d$, and (c, f) barrier width $\Delta b$ for polar (solid lines) and non-polar QWs (thick broken lines). Top row (a)-(c) is for the QWs from Fig. S1, while bottom row (d)-(f) for QWs from Fig. 2 in the manuscript.